\documentclass[aps,showpacs,floatfix,twocolumn,prl]{revtex4}

\usepackage{amsmath}
\usepackage[dvips]{graphicx}
\usepackage{epsfig}
\usepackage{tabularx}

\begin{document}

\title{Acoustic black holes in a two-dimensional ``photon fluid"}
\author{Francesco Marino}
\affiliation{Dipartimento di Fisica, Universit\`a di Firenze, INFN,
Sezione di Firenze, \\ Via Sansone 1, I-50019 Sesto Fiorentino
(FI), Italy}

\date{\today}
\begin{abstract}
Optical field fluctuations in self-defocusing media can be described in terms of sound waves in a 2D photon-fluid. It is shown that, while the background fluid couples with the usual flat metric, sound-like waves experience an effective curved spacetime determined by the physical properties of the flow.
In an optical cavity configuration, the background spacetime can be suitably controlled by the driving beam allowing the formation of acoustic ergoregions and event horizons. An experiment simulating the main features of the rotating black hole geometry is proposed.
\end{abstract}


\maketitle

\section{Introduction}
Progress in understanding critical phenomena in general relativity and quantum field theory in curved spacetime, suffers greatly from the lack of experimental feedback. For this reason, strong efforts have been made in order to find non-relativistic systems, experimentally testable in the laboratory, able to simulate gravitational spacetime geometries and cosmological solutions \cite{rev}.
The first proposal dates back to 1981, when Unruh showed that the propagation of sound waves in an inhomogeneous flowing fluid is analogous to that of a massless scalar field in curved Lorentzian spacetime \cite{unruh}.
A surface in this flow, where the normal component of the fluid velocity is inward pointing and equal to the local speed of sound behaves as the event horizon of a gravitational black hole, i.e. sound waves cannot propagate through this surface in the outward direction. This kinematical analogy implies that acoustic black holes should emit Hawking radiation \cite{hawking} in the form of a thermal bath of phonons \cite{unruh,visser1}, thus making possible the experimental verification of such phenomenon.

Following Unruh's prediction many different kinds of analogue models has been proposed \cite{vol,sch,lor} but
Bose-Einstein condensates (BEC) \cite{garay,giovanazzi} are currently the best workbenches for this kind of investigation.
The experimental implementation of acoustic black holes in such systems could be possible by
setting suitable external potentials to obtain the required background flow.
Interesting proposals have been done also in the optical domain
when Leonhardt and Piwnicki proposed to create an optical black hole \cite{leo}. The idea is feasible since
the propagation of light in a moving medium resembles many features of a curved spacetime and very recently
ultrashort pulses in optical fibers demonstrated the formation of an artificial event horizon \cite{science}.

It is the aim of this work to show that exploiting the relation between nonlinear optics and fluid dynamics
rotating acoustic black holes can be created in a self-defocusing optical cavity. Self-defocusing media possess an intensity-dependent refractive index, $n = n_0 - n_2 \vert E \vert^2$, where $n_0$ is the linear refractive index, $n_2 > 0$ is the material nonlinear coefficient and $\vert E \vert^2$ is the optical intensity. As a consequence light propagating in a self-defocusing medium induces a local negative bending of the refractive index which, in turn, affects the light beam itself. At a microscopic level this can be described in terms of an atom-mediated repulsive interaction between photons which leads to the formation of a ``photon-fluid" \cite{fr,boyce}. It will be shown that linear excitations of such fluid (sound waves) propagate
in an effective curved spacetime determined by the physical properties
of the optical flow. However, the simple propagation in a self-defocusing medium
does not allow the implementation of the optical background required to generate a stationary "black-hole" geometry.
In contrast, if the medium is placed in an optical cavity, suitable choice of the
injected beam profile leads to the formation of acoustic ergoregions and event horizons, making
possible the experimental simulation of an acoustic rotating black hole.
The possibility to control the spacetime geometry together with the fact that light signals
are much easily detectable than acoustic disturbances in fluids make the system here
proposed an interesting alternative for "analog-gravity" experiments.

\section{Acoustic Metric in self-defocusing media}

First consider the simple propagation of a monochromatic optical beam of wavelength $\lambda$, in a self-defocusing medium (without cavity). The slowly varying envelope of the optical field follows the well-known
Nonlinear Schroedinger Equation (NSE) \cite{solit}
\begin{equation}
\partial_z E = \frac{i}{2k}\nabla^{2} E - i \frac{k n_2}{n_0} E \vert E \vert^2
\label{eq1}
\end{equation}
where $z$ is the propagation direction, and $k = 2 \pi n_0 / \lambda $ is the wave number. $\nabla^{2} E$, defined with respect to the transverse coordinates $(x, y)$, accounts for diffraction, while the nonlinear term describes the self-defocusing effect.
The link with fluid dynamics becomes evident writing the complex scalar field in terms of its amplitude and phase, $E = \rho^{1/2} e^{i \phi}$. This representation converts Eq. (\ref{eq1}) into the hydrodynamic continuity and Euler equation
\begin{eqnarray}
\partial_t \rho + \nabla\cdot(\rho {\bf v}) = 0 \, \\
\partial_t \psi + \frac{1}{2} v^2 + \frac{c^2 n_2}{n_0^3}\rho
- \frac{c^2}{2 k^2 n_0^2}\frac{\nabla^2 \rho^{1/2}}{\rho^{1/2}} = 0
\end{eqnarray}
where the optical intensity $\rho$ corresponds to the fluid density, ${\bf v} = \frac{c}{k n_0}\nabla \phi \equiv \nabla \psi$ is the fluid velocity (here $c$ is the speed of light) and the propagation direction acts as time $t = \frac{n_0}{c} z$. Apart the optical coefficients, Eq. (2)-(3) are identical to those describing the density and the phase dynamics of a 2D BEC in the presence of repulsive atomic interactions \cite{dalfo} (the interaction is attractive if self-focusing nonlinearity is considered). The optical nonlinearity, corresponding to the atomic interaction, provides the bulk pressure $P = \frac{c^2 n_2 \rho^2}{2 n_0^3}$ while the last term in Eq. (3) (quantum pressure) has no analogy in classical fluid mechanics. In optics, it is a direct consequence of the wave nature of light (it arises from the diffraction term) and is significant in rapidly varying and/or low intensity regions such as dark solitons core and close to boundaries.

The evolution equation for sound waves in the photon fluid can be obtained by linearizing Eqs. (2)-(3) around a background state, since acoustic disturbances are defined as the first order fluctuations of the quantities describing the mean fluid flow. By setting $\rho = \rho_0 + \epsilon \rho_1 + O(\epsilon^2)$, and $\psi = \psi_0 + \epsilon \psi_1 + O(\epsilon^2)$ we obtain
\begin{eqnarray}
\partial_t \rho_1 + \nabla\cdot(\rho_0\nabla\psi_1 + \rho_1{\bf v_0}) = 0 \; \\
\partial_t \psi_1 + \nabla\psi_1\cdot{\bf v_0} = \nonumber
\; \\
\frac{c^2}{4 k^2 n_0^2}\left[\nabla\cdot\left(\frac{\nabla \rho_1}{\rho_0}\right) - \frac{\rho_1}{\rho_0}\nabla\cdot\left(\frac{\nabla \rho_0}{\rho_0}\right)\right]
- \frac{c^2 n_2}{n_0^3} \rho_1
\end{eqnarray}
When the quantum pressure is negligible, Eqs. (4)-(5) can be reduced to a single second order equation for the phase perturbations
\begin{eqnarray}
-\partial_t
     \left(\frac{\rho_0}{c_s^2} \;
            ( \partial_t \psi_1 + {\bf v_0} \cdot \nabla\psi_1 )
     \right)
\nonumber\\
+\nabla \cdot
     \left( \rho_0 \; \nabla\psi_1
            - \frac{\rho_0 {\bf v_0}}{c_s^2} \;
	      ( \partial_t \psi_1 + {\bf v_0} \cdot \nabla\psi_1 ) 	
     \right)
=0.
\label{wv}
\end{eqnarray}
where $c_s$ is the local speed of sound, usually defined as $c_s^{2} \equiv \frac{\partial P(\rho_0)}{\partial \rho} = \frac{c^2 n_2 \rho_0}{n_0^3}$.
Note that Eq. (\ref{wv}) has the form of a wave equation for a massless scalar field $\Delta \psi_1 \equiv
{1\over\sqrt{-g}} \partial_\mu
\left( \sqrt{-g}\; g^{\mu\nu} \; \partial_\nu\psi_1 \right) =
0$ propagating in a $(2+1)$-dimensional curved spacetime whose (covariant) metric reads
\begin{equation}
g_{\mu\nu} =
\left(\frac{\rho_0}{c_s}\right)^2 \left(%
\begin{array}{cc}
  -(c_s^2 - v_0^2)  &  -{\bf v_0^T} \\
  -{\bf v_0}  &  {\bf I} \\
\end{array}%
\right)
\end{equation}
where $g = det(g_{\mu \nu})$ and ${\bf I}$ is the two-dimensional identity matrix.
Equivalently, the line element in polar coordinates on the plane is
\begin{equation}
ds^2 = (\frac{\rho_0}{c_s})^2 [-(c_{s}^2 - v_{0}^2)dt^2 - 2 v_r dr dt - 2 v_{\theta} r d\theta dt + dr^2 + (r d\theta)^2]
\end{equation}
where $v_{0}^2 = v_{r}^2 + v_{\theta}^2$, $v_r = \partial_r \psi_0$ and $v_{\theta} = \frac{1}{r}\partial_{\theta} \psi_0$.
The acoustic metric (7), already found in BECs \cite{garay} and classical fluids \cite{unruh,visser1}, can have ergoregions and event horizons depending on the physical properties of the flow. The region in which the fluid velocity exceeds the speed of sound, i.e. $v_0^2 > c_s^2$ defines the ergosphere, where no physical objects can remain at rest relative to an inertial observer at infinity. The event horizon, instead, is defined by the surface where the radial component of the fluid velocity $v_r$ equals $c_s$.
Note however that this geometrical interpretation fails when the quantum pressure cannot be neglected, for instance when the length-scale of spatial variations of the density is smaller than the critical length $\xi = \frac{\lambda}{2 \sqrt{n_0 n_2 \rho_0}}$. This point can be clarified considering regions of nearly homogeneous background, where $\rho_1$ and $\psi_1$ can be treated as slowly varying amplitude plane waves (eikonal approximation) \cite{rev}. Then Eqs.(4)-(5) lead to a curved-space generalization of the Bogoliubov dispersion relation in a Bose gas \cite{bogo}
\begin{equation}
(\Omega - {\bf K} \cdot {\bf v_0})^2 = \frac{c^2 n_2 \rho_0}{n_0^3}K^2 + \frac{c^2}{4 k^2 n_0^2}K^4
\label{disp}
\end{equation}
where $K$ is the wave number of the sound mode and $\Omega$ its angular frequency in the laboratory frame. Assuming $v_0 = 0$ (no background velocity) it is immediate to see that modes of wavelength $\Lambda \gg \xi$ follow the standard phononic dispersion relation, $\Omega \approx c_s K$, while at high-energies ($\Lambda \ll \xi$) the quantum pressure contribution is dominant and $\Omega \approx \frac{c}{2 k n_0}K^2$. In this case the group velocity can grow without bound allowing short-wavelengths modes to escape from behind the horizon. The length $\xi$ (the analog of the "healing length" of BEC) is the scale of the violation of Lorentz invariance, usually identified in quantum gravity phenomenology with the Planck scale.

In the long-wavelength limit, where the picture of an effective metric makes sense \cite{fra}, acoustic black holes can form in presence of a suitable optical background which, however, is constrained to solve Eqs. (2)-(3). One of the most studied fluid flows to pursue "analog gravity" scenarios is the stationary vortex flow, being the hydrodynamic structure closer to rotating black holes \cite{vs}. Its optical counterpart arises from the self-trapping of a phase singularity embedded in a broad optical beam due to the counterbalanced effects of self-defocusing and diffraction \cite{solit}. The resulting structure (vortex soliton) is characterized by a dark core and a specific helical wave front, $E_0 = \rho_{0}^{1/2}(r) e^{i \psi_0}$, where $\psi_0 = m \theta$ and $\rho_0(0)=0$.
However, while such flow geometry possesses an ergosurface when $c_s(r)=v_{\theta}=\frac{c m}{k n_0 r}$ an event horizon cannot form since $v_r$ is identically zero. As remarked by Visser \cite{visser2}, provided that $c_s$ remains positive, an hydrodynamic event horizon requires the vortex to have a central sink, i.e. to possess a non-zero inward radial velocity (collapsing vortex) \cite{brev}. Unfortunately collapsing vortices appear only as transient solutions of Eq. (\ref{eq1}) and once the initial conditions has been chosen their dynamics depends only on the nonlinear wavenumber shift in the medium.
For this reason, a more controllable system in which the photon-fluid can be created and allowing the choice of the background flow profile is required. In the next section, it will be shown that both these characteristics are fulfilled in a self-defocusing optical cavity.

\section{Acoustic Horizons in an optical cavity}

In the mean field approximation, the slowly varying envelope of the intracavity field follows a damped-driven version of the NSE \cite{lugiato}
\begin{equation}
\partial_t E = \frac{i c}{2 k} \nabla^{2} E - i \omega \frac{n_2}{n_0} E \vert E \vert^2 + i \delta E - \Gamma (E - E_d)
\label{ll}
\end{equation}
where $L$ is the cavity length, $\omega$ is the pump frequency, $\delta = \omega - \omega_{cav}$ is the detuning between the pump and the cavity resonance, $E_d$ is a coherent driving field, which is proportional to the incident field, and $\Gamma = cT/2n_0L$ is the cavity decay rate, where $T$ is the mirrors transmissivity.
It is immediate to see that Eq. (\ref{ll}) is a dissipative-forced system, where the losses $-i \Gamma E$ are balanced by the coherent injected field $i \Gamma E_d$. Therefore, if $\Gamma$ is finite, the presence of the driving field "pins" the intracavity phase profile allowing the external control of the photon-fluid background, and thus the generation of "black-holes" geometries. At the same time, if $\Gamma$ is sufficiently low the photons remain trapped inside the cavity long enough so that a thermalized condition is achieved after many photon-photon interactions, thus allowing the formation of the photon fluid \cite{cav}.
\begin{figure}
\begin{center}
\includegraphics*[width=1.0\columnwidth]{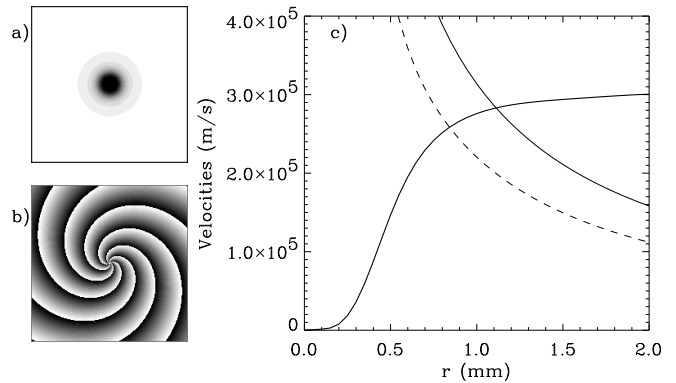}
\end{center}
\caption{a) 2D optical intensity and b) phase profile (modulo $2\pi$) obtained by numerical integration of Eq. (\ref{ll}) driven by vortex profile $\mathcal{E}_d$ with $r_0=200 \rm\mu\rm m$ and $m=6$. c) The corresponding velocities: $c_s(r)$ (bold-line), $v_0(r)$ (solid-line), $\vert v_r(r)\vert$ (dashed-line). Parameters as in the text.}
\vspace{-.3cm}
\label{figu1}
\end{figure}
As a further advantage note that, contrary to Eq. (1) where the variable $z$ acts as time, in the cavity configuration the field envelope evolves in "real time" $t$, thus allowing experimental studies of real sound modes dynamics.
The sound waves dispersion relation can be directly obtained by linearization of Eq. (\ref{ll}) around a background solution, $E_0$, describing the bulk motion of the photon fluid. By choosing the frequency of the driving field in order to operate at the maximum intracavity intensity (resonant condition) $\delta = -\frac{n_2 \omega}{n_0} \vert E_0 \vert^2$ and in the eikonal approximation we obtain
\begin{equation}
(\Omega^{'} - i \Gamma)^2 = \frac{c^2 n_2 \rho_0}{n_0^3}K^2 + \frac{c^2}{4 k^2 n_0^2}K^4
\label{disp2}
\end{equation}
where $\rho_0 = \vert E_d \vert^2$ and $\Omega^{'} = \Omega - {\bf K} \cdot {\bf v_0}$ is the sound mode frequency in the locally comoving background frame. The dispersion relation (\ref{disp2}) coincides with (\ref{disp}) except for the imaginary term $i\Gamma$ which implies dissipation. This is not surprising since Eq. (\ref{ll}) is a dissipative system having the NSE as its conservative limit ($\Gamma \rightarrow 0$). As a consequence, Eq. (\ref{disp2}) exhibits the usual breaking of the Lorentz invariance at short wavelengths while the amplitude of low-energy modes drops to $1/e$ of its initial value in correspondence of the propagation distance $L_d = c_s/\Gamma$.
However, excitations of wavelengths $\xi \ll \Lambda \ll L_d$ behave as genuine sound waves propagating with constant group velocity $c_s$ in direct analogy to scalar fields in curved spacetime. In this range indeed Eq. (\ref{disp2}) takes the form $g^{\mu\nu}K_{\mu}K_{\nu}=0$, where $K_{\mu} =(\Omega/c_s, {\bf K})$ is the covariant wave vector and $g^{\mu\nu} = (g_{\mu\nu})^{-1}$.

\subsection{Experimental Proposal}
The first step of an experimental investigation would be to study how the presence of a "black-hole" background modifies the propagation and the wavelength of sound waves. In particular, inside the the horizon, the background flow speed is larger than the local speed of sound, and so sound waves will be dragged inwards.
In order to perform this experiment we need to apply two incident coherent beams to the nonlinear cavity: a first beam resonant with the cavity forms the photon fluid and creates a suitable black-hole background, and a second narrow amplitude-modulated beam which is modulated at the desired sound wave frequency provides a local excitation in a given point of the transverse plane. This perturbation will induce sound-like waves in the photon fluid which propagate away from the point of injection in the $(x, y)$-plane.
Such waves can be detected by imaging the output mirror face and simultaneously monitoring the intensity signal at different positions of the transverse plane by fast photodetectors. The corresponding phase dynamics can be reconstructed by interference with a reference beam. The velocity and the dispersion relation of sound-waves can be obtained from this kind of measurements and the validity of the relation (\ref{disp2}) can be experimentally tested.
In order to simulate a spacetime geometry analogue to that of a rotating black hole (Kerr metric), we choose the driving field profile $\mathcal{E}_d=\sqrt{\rho_d}exp(im\theta-2i\pi\sqrt{\frac{r}{r_0}})$, where $\rho_d$ is constant. While numerous experimental methods are available to obtain these fields \cite{phase}, the most convenient approach is probably
to use a phase-only spatial light modulator. This device consist of a two-dimensional array of
individually addressable pixels, acting as electrically controllable wave plates. A computer controls the voltage applied
to each pixel allowing the possibility to encode a given phase pattern into the laser beam.
As an example, consider a $0.1$-m-long cavity with $T=2~10^{-2}$ filled with $^{85} Rb$ vapor at $80 ^{0}C$ corresponding to $10^{12} \rm atoms/cm^{3}$. The self-defocusing regime is obtained by detuning the laser to the red side of the hyperfine transition $5S_{1/2}(F = 2)-5P_{3/2}(F = 3)$ ($\lambda \sim 780$ nm) \cite{boyce,solit}. The detuning must be chosen in order to have the strongest nonlinearity compatible with an absorption lower than T. A refractive index change $\Delta n = n_2 \rho_0 = 10^{-6}$ leads to a sound speed $c_s = 3~10^{5}$m/s. For these parameters $\xi = 390 \rm\mu$m and $L_d = 10$mm, hence phononic modes are virtually undamped over the beam transverse dimensions (that can be chosen to be $\approx 4$ mm).
Fixed this value of $c_s$, a sound-like wave with $\Lambda = 1$mm can be generated by a $300$ MHz-amplitude
modulation of the perturbation beam, which can be easily achieved by an electro-optic modulator.
Numerical integration of Eq. (\ref{ll}) with the driving field $\mathcal{E}_d$ shows that, after a short transient, a stationary stable vortex profile forms. The intracavity field has a spiral equiphase surface and at the vortex centre, where the phase is singular, the intensity vanishes (see Fig. \ref{figu1}a,b).
The corresponding sound speed and flow velocity profiles reported in Fig. \ref{figu1}c define the locations of the ergosurface and event horizon. The former is found where the flow goes supersonic, i.e. $r_E \approx 1.12$mm, while the latter appears where the sound speed equals the radial component of the fluid velocity, i.e. $r_H \approx 0.84$mm.

A closed-form analytical expression for the intracavity field does not exist; however, far from the vortex core, where the long-wavelength approximation is valid, $E_0$ asymptotes to the homogeneous solution $E = \mathcal{E}_d$. In this case the acoustic line element is given by $(8)$ with $c_s = \sqrt{\frac{c^2 n_2 \rho_d}{n_0^3}}$, $v_{r}=-\frac{c \pi}{k n_0\sqrt{r_{0}r}}$ and $v_{\theta}=\frac{c m}{k n_0 r}$ and the location of the horizon and ergosurface can be explicitly estimated. The similarity with the equatorial slice of the Kerr geometry becomes clearer through the transformations of the time and the azimuthal angle coordinates in the outer region $(\xi^2 /r_0 \leq r < \infty)$
\begin{equation}
d\tilde{t} = dt + \frac{\vert v_r \vert}{(c_{s}^2 - v_r^{2})}dr ;  \\ d\tilde{\theta} = d\theta + \frac{\vert v_r \vert v_{\theta}}{r (c_{s}^2 - v_r^{2})}dr .
\end{equation}
After a rescaling of the time coordinate by $c_s$ the metric takes the form
%\begin{eqnarray}
\begin{equation}
ds^2 \propto -[1 - \frac{\xi^2}{r_0r} - (\frac{m\xi}{\pi r})^2]dt^2 + (1 - \frac{\xi^2}{r_0r})^{-1}dr^2 -2 \frac{m \xi}{\pi} d\theta dt + (r d\theta)^2 \nonumber
%ds^2 \propto -\left(1 - \frac{\xi^2}{r_0r} - \left(\frac{m \xi}{\pi %r}\right)^2\right)dt^2 + \nonumber
%\; \\
%\left(1 - \frac{\xi^2}{r_0r}\right)^{-1}dr^2 -2 \frac{m \xi}{\pi} d\theta dt + (r %d\theta)^2
%\end{eqnarray}
\end{equation}
As in the Kerr spacetime the metric has a coordinate singularity where the radial component $g_{rr}$ goes to infinity which corresponds to the event horizon, i.e. $r_H = \xi^2 /r_0$. The radius of the ergosphere is given by the vanishing of the temporal component of the metric $g_{tt}$, $r_E =\frac{1}{2}[r_H + \sqrt{r_{H}^2 + 4 m^2r_{H}r_0/\pi^2}]$. Within this surface $g_{tt}$ is negative, i.e acts like a purely spatial metric coefficient, so that no state of rest can be defined. For the values of $\xi$ and $r_0$ used before we obtain $r_E \approx 1.22$mm and $r_H \approx 0.76$mm in good agreement with numerical calculations.

\section{Discussion and Future Perspectives}

In view of analog-gravity experiments the 2D photon fluid presents important advantages respect to BEC flows. In the latters, a stationary vortex with a sink is not easily achieved experimentally as the matter must be continuously coupled out from the vortex origin, thus requiring some means of replenishment or the use of a very large condensate. On the other hand, a nonlinear optical cavity naturally support such configuration in the presence of a suitable driving field profile. The resulting steady optical vortex flow possesses two crucial properties of rotating black holes, i.e. an ergoregion and an event horizon. These features enables laboratory tests of intriguing phenomena, such as superradiance and Hawking radiation, which are insensitive to whether or not the metric satisfies the Einstein equation and robust against the high-energy breaking of Lorentz invariance \cite{visser3}.
The search of such effects will be simplified since light signals can be amplified and detected much easier than acoustic disturbances in a real fluid.
In particular, the Hawking process has a classical interpretation in terms of the power spectrum of exponentially redshifted waves emitted near the horizon which obeys the Planck distribution \cite{ch}. The conversion of positive-frequency modes to negative-frequency modes which is the basis of this effect has been recently observed \cite{rousseaux}.
Since in the photon fluid sound waves can be locally excited by injecting a second narrow beam
into the cavity and detected by fast photodetectors, their power spectrum in the presence of an event horizon can be measured.
The observation of a Planck distribution, although has nothing to do with quantum particle creation, would represent a first step towards the simulation of Hawking radiation in the laboratory.
Regarding quantum Hawking radiation, the possibility to detect it in such a system requires further investigation.

In conclusion, sound-like waves propagation in a $2D$ photon fluid has been studied as an analog for field propagation in a curved spacetime. Although the simple light-matter interaction in self-defocusing media gives rise to an effective curved spacetime, the cavity configuration provides a way to control its geometry allowing the formation of acoustic ergoregions and event horizons. The possibility to observe phenomena analogous to superradiance and Hawking radiation and the design of a feasible experimental setup are currently under investigation.

\section{Acknowledgment}
I acknowledge F. Marin and A. Ortolan for valuable discussions and the careful reading of the manuscript.


\begin{thebibliography}{99}
\bibitem{fra} Note that in the presence of an event horizon deviations from the long-wavelength limit are important even at low energies. Indeed all excitations experience a formally infinite blueshift (increase in wave number) when traced back in time towards the horizon. When $\Lambda \approx \xi$ the geometrical description in terms of the acoustic metric is no longer valid. Despite this fact, acoustic horizons still exist but at a less well-defined location and for a particular set of excitations \cite{r1,r2}.
\end{thebibliography}
\end{document}